\documentclass{egpubl}
 
\JournalPaper         
\usepackage[T1]{fontenc}
\usepackage{dfadobe}  

\usepackage{cite}  
\BibtexOrBiblatex
\electronicVersion
\PrintedOrElectronic
\ifpdf \usepackage[pdftex]{graphicx} \pdfcompresslevel=9
\else \usepackage[dvips]{graphicx} \fi

\usepackage{egweblnk}
\usepackage[title, toc, page]{appendix}

\usepackage{hyperref}
\usepackage{float}
\usepackage{cite}
\usepackage{mathrsfs}
\usepackage[table,xcdraw]{xcolor}
\usepackage{mathptmx}
\usepackage{amstext}
\usepackage{amsfonts}
\usepackage{amsmath}
\usepackage{amssymb}
\usepackage{graphicx}
\usepackage{algorithm}
\usepackage[noend]{algorithmic}
\usepackage{multirow}
\usepackage{booktabs}
\usepackage[normalem]{ulem}
\useunder{\uline}{\ul}{}

\usepackage[normalem]{ulem} 
\usepackage{figsize}

\usepackage{ifpdf}
\usepackage{wrapfig}

\usepackage{color}
\definecolor{blue}{rgb}{0,0,0.8}
\definecolor{green}{rgb}{0,0,0}
\definecolor{red}{rgb}{0.7,0,0}

\def\green{\color{green}}

\def\mat#1{\mathbf{#1}}
\def\vec#1{\mathbf{#1}}

\def\mS{$\mathbf{S}$}
\def\mD{$\mathbf{D}$}
\def\mDK{{$\mathbf{DK}$}}
\def\mDKG{{$\mathbf{DKG}$}}
\def\mDKGM{{$\mathbf{DKGM}$}}
\def\mDKGMZ{{$\mathbf{DKGMZ}$}}

\def\widehatt#1{\hat{#1}}

\newcommand{\vp} {\vec{p}}

\newcommand{\vq} {\vec{q}}

\newcommand{\vz} {\vec{z}}

\newcommand \mP	{\mat{P}}

\newcommand \mQ	{\mat{Q}}

\newcommand{\real} {\mathbb{R}}

\newcommand{\etal}{\textit{et al}.}

\title[Constructing Human Motion Manifold with Sequential Networks]%
      {Constructing Human Motion Manifold with Sequential Networks}

\author[D-K. Jang \& S-H. Lee]
{\parbox{\textwidth}{\centering Deok-Kyeong Jang
        and Sung-Hee Lee 
        }
        \\
{\parbox{\textwidth}{\centering Korea Advanced Institute of Science and Technology (KAIST)
      }
}
}
\begin{document}

\maketitle
\begin{abstract}
This paper presents a novel recurrent neural network-based method to construct a latent motion manifold that can represent a wide range of human motions in a long sequence. We introduce several new components to increase the spatial and temporal coverage in motion space while retaining the details of motion capture data. These include new regularization terms for the motion manifold, combination of two complementary decoders for predicting joint rotations and joint velocities, and the addition of the forward kinematics layer to consider both joint rotation and position errors. In addition, we propose a set of loss terms that improve the overall quality of the motion manifold from various aspects, such as the capability of reconstructing not only the motion but also the latent  manifold vector, and the naturalness of the motion through adversarial loss.
These components contribute to creating compact and versatile motion manifold that allows for creating new motions by performing random sampling and algebraic operations, such as interpolation and analogy, in the latent motion manifold.


\end{abstract}  

\section{Introduction}
\label{sec:intro}
Constructing a latent space for human motion is an important problem as it has a wide range of applications such as motion recognition, prediction, interpolation, and synthesis.
Ideal motion spaces should be compact in the sense that random sampling in the space leads to plausible motions and comprehensive so as to generate a wide range of human motions. In addition, locally linear arrangement of the semantically related hidden vectors would benefit motion synthesis, e.g., by simple algebraic operations.

However, constructing a compact and versatile motion space and extracting valid motions from it remains a challenging problem because the body parts of human body are highly correlated in general actions and the joints are constrained to satisfy the bone lengths and the range of movement. The high dimensionality of the joint space adds additional difficulty to this problem.

\begin{figure}[h]
  \centering
  \includegraphics[width=0.9\linewidth]{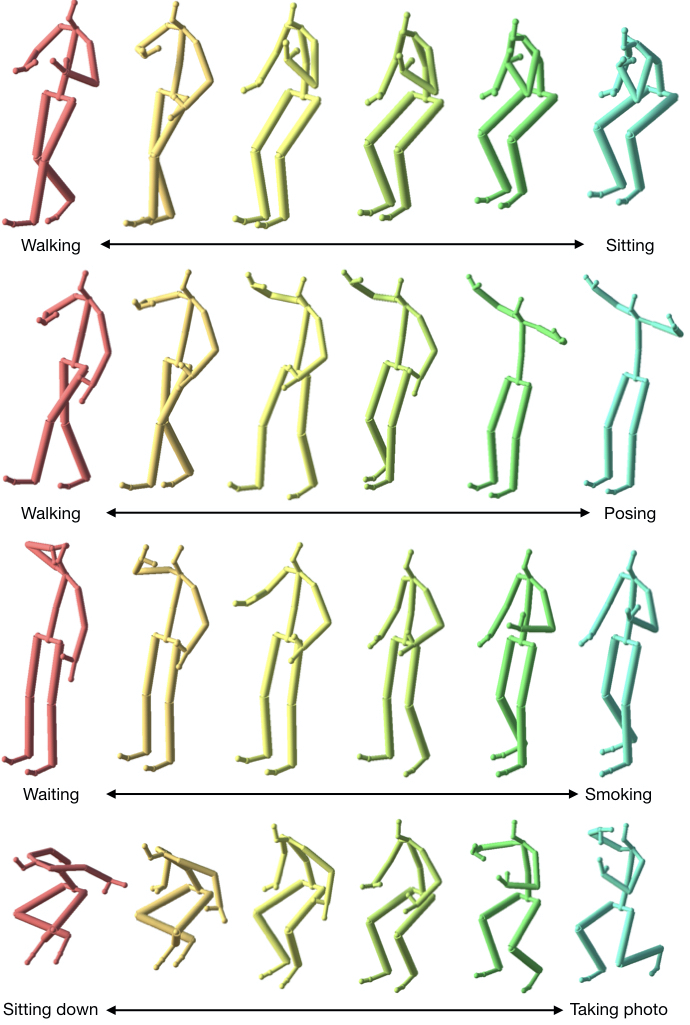}
  \caption{Examples of motion interpolation on the latent motion manifold generated by our method. The first and last columns are snapshots of two input motions, and the intermediate columns show the snapshots of four individual motions obtained by the linear interpolation on the motion manifold.}
  \label{fig:interpolation}
\end{figure}

In this paper, we present a novel framework to construct a latent motion manifold and to produce various human motions from the motion manifold.
In order to embrace the temporal characteristic of human motion, our model is based on the sequence-to-sequence model.
The unsupervised sequence-to-sequence models have been shown to be effective by previous studies on motion prediction \cite{martinez2017human, pavllo2018quaternet}. %
Based on these studies, we develop several novel technical contributions to achieve a compact yet versatile latent motion manifold and a motion generation method as follows.

First, our model is characterized by the combination of one encoder and two decoders.
Given a motion manifold vector, one decoder learns to generate the joint rotation while the other learns to output joint rotation velocities. As will be discussed later, the joint rotation decoder has the advantage of reconstructing long term motions better. 
In comparison, the joint velocity decoder has the advantage of improving the continuity of the motion. 
By complementing each other, our two decoder model shows a higher reconstruction accuracy than that of the single decoder model.

Second, unlike previous studies that deal with only either joint angles or joint positions, by adding a forward kinematics (FK) layer \cite{villegas2018neural}, our joint angle-based human representation achieves the advantage of satisfying bone-length constraints and simplifying joint limit representation. 
By additionally considering joint position computed by the FK layer while training, our method reduces the joint position error, which is visually more perceptible than the joint angle error.

Lastly, we introduce several loss functions, each of which contributes to enhancing the quality of the motion manifold in different aspects.
A reconstruction loss reduces the difference between the reconstructed motion and the input motion and thus allows the manifold to synthesize motion content and details observed in the training motion dataset. 
A regularizer loss improves the distribution quality of the motion manifold and thus enables random sampling and interpolation on the manifold.
In addition, an adversarial loss increases the naturalness of the motions generated from the motion manifold.

In this paper we show that, based on these technical contributions, our method allows for various practical applications such as random generation of motions, motion interpolation, motion denoising and motion analogy as will be shown in Sec.~\ref{sec:result}. The capability of our method is demonstrated by the comparison with other approaches, such as the seq2seq model~\cite{martinez2017human} and the convolution model~\cite{holden2015learning, holden2016deep}.

The remaining part of this paper proceeds as follows: After reviewing previous studies related to our work in Sec.~\ref{sec:related}, we present our method and loss function in detail in Sec.~\ref{sec:method}. 
Sections~\ref{sec:data} detail the data pre-processing and Sec.~\ref{sec:result} reports a number of experiments performed to verify the effectiveness of our method. 
Section~\ref{sec:conclusion} discusses the limitations of our work, future research directions, and concludes the paper.
Our code and networks are available at {\small \url{https://github.com/DK-Jang/human_motion_manifold}}.

\section{Related work}
\label{sec:related}
Researcher have developed several methods to construct motion manifold to generate natural human motions, but compared with studies on manifold learning for other data such as image, research on motion data is scarce.
Linear methods such as PCA can model human motion in only a local region. 
Chai \etal \cite{chai2005performance} apply local PCA to produce a motion manifold that includes a certain range of human motion, and apply it for synthesizing movements from low dimensional inputs such as the position of end effectors. 
Lawrence \cite{lawrence2004gaussian} use Gaussian Process Latent Variable Model (GPLVM) to find a low dimensional latent space for high dimensional motion data.
Taylor \etal \cite{taylor2007modeling} propose a modified Restricted Boltzmann Machine that is able to deal with the temporal coherency of the motion data. Lee \etal \cite{lee2010motion} propose motion fields method, a novel representation of motion data, which allows for creating human motion responsive to arbitrary external disturbances.
Recently, with the development of deep learning technology, a method of constructing a motion manifold by using Convolutional Neural Network (CNN)-based encoder was introduced by Holden \etal \cite{holden2015learning, holden2016deep}. 
Butepage \etal \cite{butepage2017deep} compare a number of deep learning frameworks for modeling human motion data.

Our method for constructing motion manifold is based on previous studies on sequence learning for motion to predict the joint position sequences of a 3D human body given past motions.
Martinez \etal \cite{martinez2017human} develop a novel sequence-to-sequence encoder-decoder model that predicts human motion given a short duration of past motion. The presented result is impressive but has a few limitations that sometimes implausible motions such as foot sliding are generated and the initial pose of the predicted motion is somewhat discontinuous from the input motion.

Pavllo \etal \cite{pavllo2018quaternet} selectively use a joint rotation-based loss for short term prediction and a joint position-based loss for long term prediction. The latter includes forward kinematics to compute the joint positions. However, the basic sequence-to-sequence model can only predict short term motions and has limitations in predicting non-trivial, long term motions. In addition, a loss function that minimizes only the prediction error does not guarantee to construct compact and versatile motion manifold. Our method solves these problems by jointly considering joint rotation and position errors in the loss function and by adding regularization to the motion manifold. 

In a broader perspective, our work is related with the studies on recognizing and generating human motion, which 
remains a challenging research topic due to the high dimensionality and dynamic nature of the human motion.
Wu and Shao \cite{wu2014leveraging} propose a hierarchical dynamic framework that extracts top-level skeletal joint features and uses the learned representation to infer the probability of emissions to infer motion sequences. Du \etal ~\cite{du2015hierarchical} and Wang \etal ~\cite{wang2017modeling} use recurrent neural network (RNN) to model temporal motion sequences and propose hierarchical structure for action recognition.
With regard to motion synthesis, Mittelman \etal \cite{mittelman2014structured} propose a new class of  Recurrent Temporal Restricted Boltzmann Machine (RTRBM). The structured RTRBM explicitly graphs to model the dependency structure to improve the quality of motion synthesis.
Fragkiadaki \etal \cite{fragkiadaki2015recurrent} propose the Encoder-Recurrent-Decoder (ERD) that combines representation learning with learning temporal dynamics for recognition and prediction of human body pose in videos and motion capture. 
Jain \etal \cite{jain2016structural} propose structural RNN for combining the power of high-level spatio-temporal graphs.

\section{Method}
\label{sec:method}
This section details our framework. 
After defining notations used in this paper, we explain the  structure of the network and the design of the loss function for training.

\subsection{Representation and notations}
\label{subsec:rep}
We denote the human motion set by $\mathcal{Q}$ and corresponding random variable by $\mQ$. 
A motion with a time range of $[t,t+\Delta t -1]$ is written as $\mQ_{t:(t+\Delta t -1)} = [\vq_t, \ldots, \vq_{t+\Delta t -1}]$, where $\vq_t$ denotes the pose  at time $t$. A pose is represented with a set of joint angles written in the exponential coordinates, i.e., $\vq_t = [q_{i,x}^t, q_{i,y}^t, q_{i,z}^t]_{i=1}^{n_{joint}}$ where $(q_{i,x}^t, q_{i,y}^t, q_{i,z}^t)$ are the three components of the exponential coordinates and $n_{joint}$ is the number of joints. 
Therefore, the dimension of a human motion is $\mathcal{Q} \in \real^{\Delta t \times n_{joint} \times 3}$.
Lastly, $\vp_t$ is the pose represented with the joint positions at time $t$ corresponding to $\vq_t$, and $\mP_{t:(t+\Delta t -1)}=[\vp_t, \ldots, \vp_{t+\Delta t -1}]$.
$\mP$ is also a random variable of motion set $\mathcal{Q}$.

\subsection{Motion manifold with sequential networks}
\label{subsec:motionmanifold}
\begin{figure*}[th]
  \centering
  \includegraphics[width=0.99\textwidth]{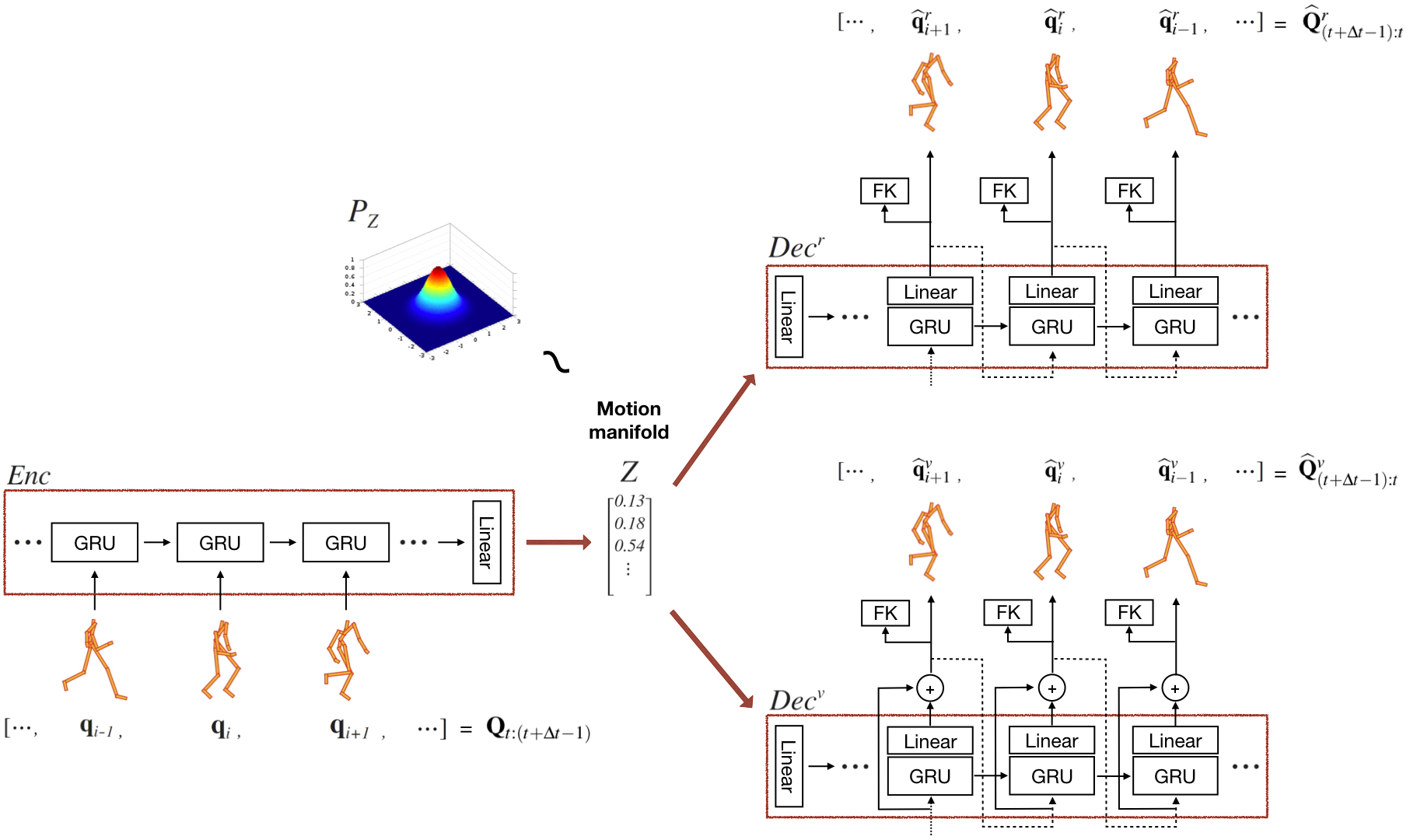}
  \caption{Structure of our sequential networks for constructing the motion manifold.}
  \label{fig:overview}
\end{figure*}

We construct a motion manifold in an end-to-end unsupervised way using a network of sequential networks, with an objective to minimize the difference between the ground truth motion space distribution and the reconstructed motion space distribution extracted from the latent motion manifold. 
To this end, we develop a sequential model that consists of the RNN with Gated Recurrent Unit (GRU). Our model has a sequence-to-sequence structure \cite{martinez2017human}, which is often used in machine translation. This RNN structure is effective for maintaining the temporal coherency in motion, and {\green it is trained to generate a fixed length of motion (150 frames) in our study.}   
As shown in Fig. \ref{fig:overview}, our model includes the combination of one encoder and two decoders with a regularizer.
The encoder takes the source motion as an input and maps it to the latent motion space.
The regularizer encourages the encoded motion distribution to approximate some prior distribution.
The two decoders are designed to map the latent motion space to joint angles and joint velocities, respectively.
Details of our model are given next.

\subsubsection{Encoder}
\label{subsubsec:enc}
The encoder consists of a GRU and one linear layer, and Fig. \ref{fig:overview} shows the unrolled schematic diagram of the encoder.
The $\Delta t$ poses $[\vq_t, \ldots, \vq_{t+\Delta t -1}]$ of a motion are input to the GRU sequentially.
The GRU encodes the current frame while being conditioned by the previous frames with their hidden representation. Specifically, the pose $\vq_i$ in the $i$-th frame is encoded as follows:
\begin{equation} \label{eq:en}
    \begin{aligned}
    h_i^{Enc} = \text{GRU}_{W_{Enc}}(h_{i-1}^{Enc}, \, \vq_i),
    \end{aligned}
\end{equation}
where $h_{i}$ is the hidden state at frame $i$, and $W_{Enc} \in \mathbb{R}^{\text{3}n_{\text{joint}} \times d_h}$ are the training parameters with $d_h$ being the hidden dimension of the GRU.
After the final pose of the input motion is read, one linear layer of parameter $W_{\text{c}} \in \mathbb{R}^{d_h \times d_m}$ receives $h_{t+\Delta t -1}$ and compresses it to produce the $d_m$-dimensional code $Z \in \mathcal{Z}$ where $\mathcal{Z}$ denotes the motion manifold. 
It is worth mentioning that this compression brings the benefit of denoising input data.
Now the encoder mapping $Enc: \, \mathcal{Q} \rightarrow \mathcal{Z}$ is completed.

\subsubsection{Latent motion manifold with the Wasserstein regularizer}
\label{subsubsec:reg}
We adopt the Wasserstein regularizer for matching the distribution $E_Z:=\mathbb{E}_{P_{\mQ}}[E(Z \mid \mQ)]$ of the motion manifold to the desired prior distribution $P_Z$.
Unlike the variational auto-encoder \cite{rezende2014stochastic}, the sequential networks trained with the Wasserstein regularizer allows non-random encoders to deterministically map inputs to the latent codes, and thus it helps randomly sampled or interpolated points in the motion manifold correspond to plausible motions.
Refer to \cite{tolstikhin2017wasserstein} for more details about the Wasserstein regularizer.

\subsubsection{Decoder with joint rotation and joint velocity}
\label{subsubsec:dec}
Our decoder model consists of two kinds: One decoder learns the joint rotation and the other learns joint rotational velocity as shown in Fig. \ref{fig:overview}. 
Both decoders are based on the GRU while the connection structures of the two are different. 
Unlike the rotation decoder, the velocity decoder adds a residual connection between the input and the output to construct joint rotation. 
Each decoder then generates the reconstructed joint angle sequence in reverse temporal order as suggested by \cite{srivastava2015unsupervised}.
The decoders are trained simultaneously with backpropagation.

This dual decoder model is based on the idea of \cite{srivastava2015unsupervised}.
By combining the two decoders, we can alleviate the limitations of individual decoder models.
The rotation decoder shows strength when reconstructing long term motions because it learns joint angle itself.
Conversely, it may cause pose discontinuity between frames.
The velocity decoder has the advantage of reconstructing continuous human motion as it outputs difference between consecutive rotations, which is usually small and easier to learn. However, training velocities tends to be unstable in a long-term sequence because the longer the motion is, the more error is accumulated. 
As our two decoders have contrasting strengths and weaknesses, when combined, they complement each other in synergy.

Unlike previous studies about motion prediction, recognition and manifold ~\cite{butepage2017deep,martinez2017human,holden2015learning,fragkiadaki2015recurrent,pavllo2018quaternet} in which either only the joint rotations or the joint positions are used, our model considers both the joint rotations and positions in the motion reconstruction loss term, $L_R$ (See Eq.~\ref{eq:motion_recon_loss}). 
Loss with joint angles has the advantage of preventing errors such as inconsistent bone length or deviation from human motion range, and thus learning with joint angle loss can generate plausible motions. 
However, rotation prediction is often paired with a loss that averages errors over joints by giving each joint the same weight. 
The ignorance of varying influence of different joints on the reconstructed motion can yield large errors in the important joints and degrade the quality of the generated poses.

The joint position loss minimizes the averaged position errors over 3D points, which better reflects perceptual differences between poses.
To combine both joint rotations and positions in the motion reconstruction loss $L_R$, we add a forward kinematics (FK) layer that computes the joint positions from the joint rotations. 
This allows for calculating the loss between the joint positions of the target motion and the reconstruction motion.
The FK module is valid for network training because its output is differentiable with respect to joint rotation.

Finally, our method reconstructs the motion in the reverse order of the input sequence.
Reversing the target sequence has an advantage in learning in that the first output frame of the decoder needs only to match the last frame input of the encoder, which allows for a continuous transition of hidden space vectors from the encoder to the decoders. {\green Refer to \cite{srivastava2015unsupervised} for a theoretical background on this approach.}
Details of our decoder are explained next.

\paragraph*{Joint Rotation Decoder}
\label{par:dec_rot}
The unfolded schematic diagram of the joint rotation decoder is shown in the upper row in Fig. \ref{fig:overview}.
It first transforms an element of the motion manifold $z \in Z$ to a $d_h$-dimensional hidden space vector with a linear layer of parameter $W_e^r \in \mathbb{R}^{d_m \times d_h}$. 
Then, conditioned by the hidden space vector representing the future frames, the GRU and a linear layer outputs the reconstructed pose $\widehat{\vq}_i^r$ at the $i$-th frame given its next pose $\widehat{\vq}_{i+1}^r$:
\begin{align}
     h_i^{{Dec}^r} &= \text{GRU}_{W_{{Dec}^r}}(h_{i+1}^{{Dec}^r}, {\green \widehat{\vq}_{i+1}^r} ), \label{eq:de_rot}
     \\
     \widehat{\vq}_{i}^r &= W_o^{r \, T} h_i^{{Dec}^r}, \label{eq:out_rot}
\end{align}
where $W_{{Dec}^r} \in \mathbb{R}^{\text{3}n_{\text{joint}} \times d_h }$ is learning parameter of the GRU and $W_o^r \in \mathbb{R}^{d_h \times \text{3}n_{joint}}$ is the parameter of the linear layer.

Note that, as mentioned earlier, the decoder uses the reversed input motion as the target motion, so the reconstruction is performed in the order of $\widehat{\mQ}_{(t+\Delta t -1):t} = [\widehatt{\vq}_{t+\Delta t -1}, \ldots, \widehat{\vq}_t]$.
Unlike the encoder, the decoder uses the reconstructed result of the previous frame as the input \cite{martinez2017human, li2017auto}.
This is equivalent to the noise scheduling \cite{bengio2015scheduled} without parameter tuning for long term reconstruction, and it also helps prevent the overfitting.
The initial input $\widehat{\vq}_{t+\Delta t}^r$ to the GRU is set zero because there is no reconstruction result of the previous frame. 
The reconstructed joint rotations are used to calculate the angle loss with respect to the target motion, and are also used to calculate the position $\widehat{\vp}_{i}^r$ through the FK layer.
\begin{equation} \label{eq:out_rot_fk}
    \begin{aligned}
        \widehat{\vp}_{i}^r = \, \text{Forward Kinematics} \, (\widehat{\vq}_{i}^r)
    \end{aligned}
\end{equation}
After the last pose $\widehat{\vq}_t$ is generated, the joint decoder mapping $Dec^r: \, \mathcal{Z} \rightarrow \mathcal{Q}$ is completed.

\paragraph*{Joint Velocity Decoder}
\label{par:dec_vel}
The joint velocity decoder has the similar structure to the joint rotation decoder.
The main difference is that it has a residual connection to generate $\widehat{\vq}_{i}^v$.
\begin{align} 
    h_i^{\text{dec}^v} &= \text{GRU}_{W_{\text{Dec}^v}}(h_{i+1}^{\text{Dec}^v}, \,  {\green \widehat{\vq}_{i+1}^v} ), \label{eq:de_vel} \\
    \widehat{\vq}_{i}^v &= W_o^{v \, T} h_i^{\text{Dec}^v} + \, \widehat{\vq}_{i+1}^v, \label{eq:de_vel_output}\\
    \widehat{\vp}_{i}^v &= \, \text{Forward Kinematics} \, (\widehat{\vq}_{i}^v), \label{eq:out_vel_fk}
\end{align}
where $W_{\text{Dec}^v} \in \mathbb{R}^{\text{3}n_{\text{joint}} \times d_h}$ and $W_o^{v}$ are the learning parameters.
This residual network learns the difference between the current frame pose $\widehat{\vq}_{i}^v$ and the previous frame pose $\widehat{\vq}_{i+1}^v$.
Therefore, the model predicts the angle difference or velocity and integrates it over time.
After the last pose is generated, the joint velocity decoder mapping $Dec^v: \, \mathcal{Z} \rightarrow \mathcal{Q}$ is completed.

\subsection{Training the motion manifold}
\label{subsec:train}
\begin{figure*}[th]
  \centering
  \includegraphics[width=0.85\textwidth]{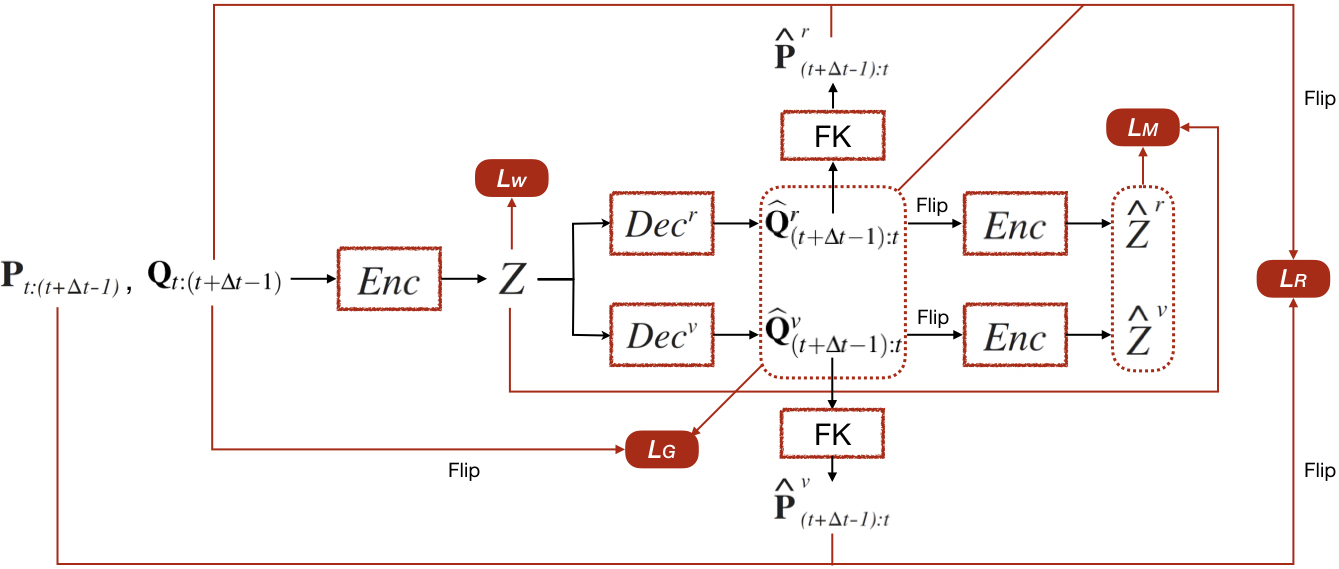}
  \caption{Each loss term is evaluated from the data processed in the network pipeline shown with black arrows. Red arrows indicate the data used for the individual loss terms.}
  \label{fig:loss_overview}
\end{figure*}

We model a number of loss functions, each of which contributes to enhancing the quality of the motion generated from the motion manifold from different perspectives.
To reduce the reconstruction loss, we employ two kinds of loss functions: Motion reconstruction loss $L_R$ that encourages a motion to be reconstructed after going through the encoder and decoder, and manifold reconstruction loss $L_M$ that helps a latent vector be reconstructed after going through the decoder and encoder. 
In addition, we include Wasserstein loss $L_W$ that penalizes the discrepancy between $P_Z$ and the distribution $E_Z$ induced by the encoder, and an adversarial loss $L_G$ to achieve more natural motions from the motion manifold.
Figure~\ref{fig:loss_overview} shows overview of our loss functions. 

\paragraph*{Motion reconstruction loss}
\label{par:motion_recon_loss}
The motion reconstruction loss penalizes the difference between the motion and the reconstructed motion, which is obtained by encoding the motion followed by decoding it.
Specifically, we measure the discrepancy of both the joint rotation angle $\vq$  and the joint position $\vp$ as follows:
\begin{align}
L_R &= L_{ang} + w_p L_{pos} \label{eq:motion_recon_loss} \\
L_{ang} &= \sum_i^{n_{joint}} \parallel \widehat{\vq}_{i}^r - \vq_i \parallel + \parallel \widehat{\vq}_{i}^v - \vq_i \parallel \\
L_{pos} &= \sum_i^{n_{joint}} \parallel \widehat{\vp}_{i}^r - \vp_i \parallel + \parallel \widehat{\vp}_{i}^v - \vp_i \parallel, 
\end{align}
where $\parallel \cdot \parallel$ is the Euclidean norm and $w_p$ (= 5 in our experiment) is the weight of the position error. 

\paragraph*{Manifold reconstruction loss}
\label{par:manifold_recon_loss}
A latent code sampled from the latent distribution should be reconstructed after decoding and encoding.
Manifold reconstruction loss encourages this reciprocal mapping between the motions and the manifold space. 
To this end, we apply $L_{1}$ loss similar to \cite{lee2018diverse}.
We draw a motion manifold vector $Z$ from the encoded motion sequences and reconstruct it with $\widehat{Z}^r = Enc(Dec^r(Z))$ and $\widehat{Z}^v = Enc(Dec^v(Z))$, where $Z = Enc(\mQ_{t:(t+\Delta t -1)})$.
\begin{equation} \label{eq:manifold_recon_loss}
    L_M = \, \parallel \widehat{Z}^r - Z \parallel_1 \, + \parallel \widehat{Z}^v - Z \parallel_1
\end{equation}

\paragraph*{Wasserstein regularizer loss}
\label{par:wa_loss}
In order to make the manifold space have a particular desired prior distribution so that we can efficiently sample from the distribution, we use the Wasserstein regularizer that penalizes deviation of the distribution $E_Z$ of the latent manifold from the desired prior distribution $P_Z$. 
\begin{equation} \label{eq:MMD}
    L_W = \text{MMD}_k(P_Z,E_Z),
\end{equation}
where $P_Z(Z) = \mathcal{N}(Z; \, \mathbf{0}, \, \sigma_z^2 \cdot \mathbf{I_d})$ is modeled as the multivariate normal distribution with $\sigma_z^2$ being decided through validation. We use the maximum mean discrepancy $\text{MMD}_k$ to measure the divergence between two distributions with the inverse multi-quadratics kernel $k(x,y)=C/(C+\parallel x-y \parallel_2^2)$ with $C=2Z_{\text{dim}}\sigma_z^2$. We set $\sigma_z^2=1$ and the dimension of motion manifold space $Z_{\text{dim}}=64$.


\paragraph*{Adversarial loss}
\label{par:ad_loss}
Finally, we employ the least squares generative adversarial network (LSGAN) to match the distribution of generated motion to the real motion data distribution, i.e., to promote motions generated by our model to be indistinguishable from real motions.
\begin{equation} \label{eq:dis_loss}
    \begin{aligned}
    L_D =&\frac{1}{2} \sum_{\widehat{\mQ}_{t:(t+\Delta t -1)}} \left[D(\widehat{\mQ}_{t:(t+\Delta t -1)})-0 \right]^2 \,\, + \\ &\frac{1}{2} \sum_{\mQ_{t:(t+\Delta t -1)}} \left[D(\mQ_{t:(t+\Delta t -1)})-1 \right]^2
    \end{aligned}
\end{equation}
\begin{equation} \label{eq:GAN_loss}
    L_G =\frac{1}{2} \sum_{\widehat{\mQ}_{t:(t+\Delta t -1)}} \left[D(\widehat{\mQ}_{t:(t+\Delta t -1)})-1 \right]^2 \,\,\,\,\,\,
\end{equation}
where the discriminator $D$ tries to distinguish between the reconstructed motions and the real motions.
The discriminator is then used to help our decoder generate realistic motions.

\paragraph*{Total loss}
\label{par:total_loss}
We jointly train the encoder, joint rotation decoder, joint velocity decoder and discriminator to optimize the total objective function, which is a weighted sum of the reconstruction loss, Wasserstein regularizer loss and adversarial loss.
The total objective function of manifold network is:
\begin{equation} \label{eq:GAN_loss}
    \begin{aligned}
    \min_{Enc,\, Dec^r,\, Dec^v} & \,\, L(Enc, Dec^r, Dec^v) \\
    & = L_R + \lambda_M \,\, L_M +\lambda_W \,\, L_W + \lambda_G \,\, L_G
    \end{aligned}
\end{equation}
and the discriminator loss is:
\begin{equation} \label{eq:GAN_loss}
    \min_{D} \,\, L(D) = \lambda_G \,\, L_D,
\end{equation}
where weighting parameters $\lambda_M$, $\lambda_W$ and $\lambda_G$ are $0.001$, $0.1$, and $0.001$ determined through validation.


\section{Data pre-processing}
\label{sec:data}
We tested our method with H3.6M dataset.
Every motion in the dataset has the same skeletal structure.
All the poses are represented with the position and orientation of the root and the joint rotations expressed with the exponential coordinates.
For the training, motion clips of 150 frames are randomly selected from the input motion sequence and used to learn a motion manifold.
The root position in the transverse plane is removed and other data are normalized for better performance.
We will explain how motion dataset is processed.

\paragraph*{H3.6M dataset}
H3.6M dataset \cite{h36m_pami} consists of 15 activities such as walking, smoking, discussion, taking pictures, and phoning performed by 7 subjects.
We reduce 32 joints in the original data to 17 joints by removing redundant joints as done by \cite{martinez2017human}, and configured all data to have a frame rate of 25 Hz.
Therefore, 150 frames motion applied to our model cover 6 seconds.
The activities of subject S5 were used as the test data and those of the remaining subjects S1, S6, S7, S8, S9 and S11 were used as the training data.
{\green Some motion data contain noises such as joint popping, but was used without noise removal.}


\section{Experimental Results}
\label{sec:result}
We perform several experiments to evaluate the performance of our method.
First, we compare the reconstruction accuracy of the proposed model with its own variations with some components ablated as well as the sequence-to-sequence model proposed by \cite{martinez2017human}.
Next, we test random sampling, motion interpolation via motion manifold, and motion denoising, followed by an experiment for motion analogies. {\green For these tests, we use the joint rotation decoder to generate motions.} We qualitatively compare the result of motion interpolation and motion analogies with that of \cite{holden2015learning} \footnote{\cite{holden2015learning} is not compared with ours with respect to the reconstruction quality as it deals only with joint positions and not joint angles.}. All experiments were conducted with test sets not included in the training set. The supplemental video shows the resulting motions from the experiments.

\subsection{Motion and manifold reconstruction}
\label{subsec:result_recon}

\begin{table*}[t]
\centering
\resizebox{0.9\textwidth}{!}{%
\begin{tabular}{@{}ccccccccccccc@{}}
\toprule
\multicolumn{2}{c}{}                        & \multicolumn{2}{c}{1.2s}        & \multicolumn{2}{c}{2.4s}        & \multicolumn{2}{c}{3.6s}        & \multicolumn{2}{c}{4.8s}        & \multicolumn{2}{c}{6.0s}                                                    &                            \\ \cmidrule(lr){3-12}
\multicolumn{2}{c}{\multirow{-2}{*}{Model}} & $E_r$             & $E_p$             & $E_r$             & $E_p$             & $E_r$             & $E_p$             & $E_r$             & $E_p$             & $E_r$             & $E_p$                                                         & \multirow{-2}{*}{$E_z$} \\ \midrule
                                & rot       & 0.889    & 0.957          & 0.971          & 0.978          & 0.990          & 1.040          & 1.097          & 1.078          & 1.195          & \multicolumn{1}{c|}{1.181}                                 & 0.317             \\
\multirow{-2}{*}{{\mS}}           & vel       & -              & -              & -              & -              & -              & -              & -              & -              & -              & \multicolumn{1}{c|}{-}                                     & -                          \\
                                & rot       & \textbf{0.823}          & 0.855          & {\ul 0.868}    & 0.923          & 0.925          & 0.999          & 1.039          & 1.032          & 1.164          & \multicolumn{1}{c|}{1.167}                                 & 0.264                      \\
\multirow{-2}{*}{{\mD}}       & vel       & {\ul 0.856}              & 0.889              & \textbf{0.843}              & 0.889              & {\ul 0.877}              & 0.961              & 1.008              & 1.081              & 1.127              & \multicolumn{1}{c|}{1.212}                                     & 0.259                          \\
                                & rot       & 1.020          & 0.561          & 1.099          & 0.682          & 1.110          & 0.706          & 1.195          & 0.761    & 1.261          & \multicolumn{1}{c|}{0.822}                                 & 0.196                      \\
\multirow{-2}{*}{{\mDK}}           & vel       & 1.347          & 0.600    & 1.353          & 0.698 & 1.323    & 0.723 & 1.382          & 0.756 & 1.391    & \multicolumn{1}{c|}{{\color[HTML]{000000} 0.809}} & 0.288                    \\
                                & rot       & 0.986          & {\ul 0.549}          & 1.077          & \textbf{0.657}          & 1.094          & {\ul 0.679}          & 1.180          & 0.726          & 1.251          & \multicolumn{1}{c|}{0.810}                                 & 0.188                      \\
\multirow{-2}{*}{{\mDKG}}           & vel       & 1.343          & 0.589 & 1.345          & 0.682    & 1.332          & 0.702    & 1.405    & 0.765          & 1.415         & \multicolumn{1}{c|}{ 0.834}                           & 0.307                      \\
                                & rot       & 0.997          & \textbf{0.541}          & 1.066         & {\ul 0.659}          & 1.084          & \textbf{0.668}          & 1.162          & \textbf{0.696}          & 1.258          & \multicolumn{1}{c|}{\textbf{0.780}}                                 & 0.182                      \\
\multirow{-2}{*}{{\mDKGM} (ours)}         & vel       & 1.356         & 0.590          & 1.381          & 0.673          & 1.338          & 0.694          & 1.400          & 0.735          & 1.406          & \multicolumn{1}{c|}{0.792}                                 & 0.293                \\
                                & rot       & 0.906          & 0.629          & 0.909          & 0.730          & 0.886          & 0.724          & {\ul 0.954}          & 0.754          & {\ul 1.053}          & \multicolumn{1}{c|}{{\ul 0.788}}                                 & {\ul 0.164}                      \\
\multirow{-2}{*}{{\mDKGMZ}}           & vel       & 0.877          & 0.635 & 0.883          & 0.703    & \textbf{0.848}          & 0.689    & \textbf{0.916}    & {\ul 0.706}          & \textbf{1.030}          & \multicolumn{1}{c|}{0.815}                           & \textbf{0.157}                      \\
                                & rot       & -              & -              & -              & -              & -              & -              & -              & -              & -              & \multicolumn{1}{c|}{-}                                     & -                          \\
\multirow{-2}{*}{Seq2seq}       & vel       & 0.875 & 0.863          & 0.870  & 0.954          & 0.891 & 1.059          & 1.039 & 1.177          & 1.154 & \multicolumn{1}{c|}{1.258}                                 & 0.216                      \\ \bottomrule
\end{tabular}
}
\caption{{\green Reconstruction errors of joint angles ($E_r$) and joint positions ($E_p$) at sample time frames, and the reconstruction error of the manifold vector ($E_z$). The error is measured with respect to the general actions (all the actions in the DB) in H3.6M dataset.}}
\label{table:H3.6M_general}
\end{table*}

We assess the {accuracy} of the reconstructed motion $\widehat{\mQ}$ with respect to the input motion $\mQ$,
as well as the {accuracy} of the reconstructed motion manifold vector $\widehat{\vz}$ with respect to the motion manifold vector $\vz$ obtained by encoding a motion. The results are provided in Table~\ref{table:H3.6M_general}.
Generally, the reconstruction {accuracy} and the data generation quality of a manifold conflict with each other to some degree. As our purpose is to achieve a motion manifold that supports not only the motion reconstruction but also motion generation, it is important to strike a balance among various performance measures, and our method should not be evaluated only by the reconstruction {accuracy}. This trade off will be discussed in Sec. \ref{subsec:tradeoff}. 

The sequence-to-sequence model (Seq2seq) compared with ours is based on \cite{martinez2017human}. The only difference is that a fully connected layer of 64 dimension is implemented between the encoder and the decoder to construct a motion manifold. 

{\green 
For ablation study, we prepare a set of variations of our model. 
The most basic model, denoted {\mS}, has only joint rotation decoder  with reconstruction and Wasserstein regularizer losses, without the FK layer in the network. Next model {\mD} is the dual decoder model by adding the velocity decoder. From the dual model, we make variations by incrementally accumulating FK layer ({\mDK}), adversarial loss ({\mDKG}), manifold reconstruction loss ({\mDKGM}, our method).
The last variation {\mDKGMZ} is made by concatenating the manifold vector to the decoder input, i.e., $\left[ \widehat{\vq}_{i+1}, \, Z \right]$ is used instead of $\widehat{\vq}_{i+1}$ in Eqs. \ref{eq:out_rot} and \ref{eq:de_vel}. The idea of this last variation is to prevent the decoder from forgetting the motion manifold vector. 
%
}
All variations 
have the same network weight dimensions and hyper-parameters as our model. Supplemental material includes details of implementing the compared models. All models are trained with datasets that include all action categories.

The {accuracy} of the motion reconstruction is evaluated for both the joint rotation decoder ($Dec^r$) and the joint velocity decoder ($Dec^v$). 
Both the Euclidean distances of joint angle errors ($L_{ang}$, also denoted as $E_r$) and joint position errors ($L_{pos}$ or $E_p$)  are {\green used for each decoder for the reconstruction loss}.
As for the reconstruction quality of the motion manifold vector, we measure the $L_1$-norm ($E_z$) of the difference between the motion manifold vector $\vz$ obtained by encoding a motion sequence and the reconstructed vector $\widehat{\vz}^r$ obtained by sequentially decoding $\vz$ and encoding it.

\begin{figure}[h]
  \centering
  \includegraphics[width=0.99\linewidth]{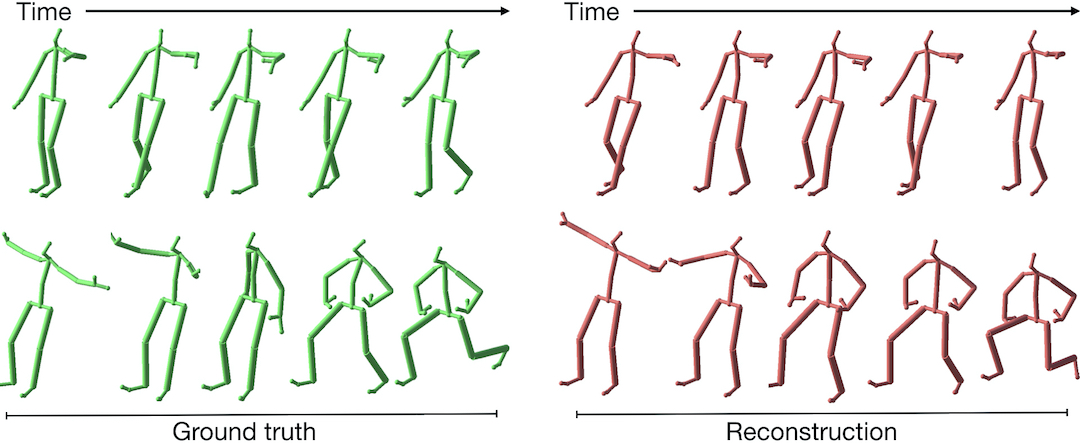}
  \caption{Ground truth motions (green) and reconstruction results (coral) of our method from H3.6M dataset.}
  \label{fig:recon}
\end{figure}

Table~\ref{table:H3.6M_general} shows the reconstruction errors of our method and others for the datasets containing all action categories (15 actions in H3.6M dataset).
The reported errors are the average of 30 motions randomly selected from a test dataset. 
A total of 150 frames are divided into 5 intervals, and errors ($E_r$, $E_p$) are measured for each interval to investigate the temporal characteristic. 
The lowest and the next lowest errors are marked in bold and with underline, respectively.

{\green 
We first compare with respect to $E_r$ and $E_p$ errors. 
Comparing {\mS} and {\mD}, the latter has lower $E_r$ and $E_p$ errors, which suggests that the joint rotation and velocity decoders complement with each other to reduce the errors.
Comparing {\mD} and {\mDK}, the latter reduces $E_p$ error significantly while only mildly sacrificing $E_r$ error. 
{\mDKG} has lower $E_r$ and $E_p$ errors than {\mDK}, but higher errors than {\mD} and {\mS}. 
This shows that adversarial loss slightly reduces reconstruction error. However, it turns out that the adversarial loss helps reconstruct the original behaviors, as will be discussed in Sec.~\ref{subsec:ZAd}.
Examining the error of {\mDKGM} and {\mDKGMZ}, we can see that adding manifold reconstruction loss does not significantly affect the reconstruction errors while explicitly feeding the manifold vector to the decoder helps reduce the errors. 

Next, we examine manifold reconstruction error, $E_z$ ({\green = $L_M$}).
Comparing {\mD} and {\mS}, it is remarkable that {\mD} reduces $E_z$ error even without any manifold-related loss term.
However, adding FK layer to reduce joint position error slightly increases $E_z$ for the velocity decoder while it is decreased for the rotation decoder. Comparing {\mDK} and {\mDKG}, we can see that adversarial loss has negligible effect to the manifold reconstruction error.
Subsequently, {\mDKGM} reduces $E_z$ slightly by adding the manifold reconstruction error, and {\mDKGMZ} achieves the lowest $E_z$ error by explicitly feeding the manifold vector to the decoder.

{\green Seq2seq~\cite{martinez2017human} shows less $E_r$ than our model, but $E_p$ is higher. In addition, our model shows better $E_z$ errors with respect to rotation decoder.}
Figure~\ref{fig:recon} visualizes the reconstruction results with our model over time in comparison with the ground truth input motion.
}

\subsubsection{Trade off between joint angle, joint position and motion manifold}
\label{subsec:tradeoff}
\begin{figure*}[th]
\centering
  \subfigure[$L_{ang}$ with respect to $\lambda_W$]
  {\includegraphics[width=0.33\linewidth]{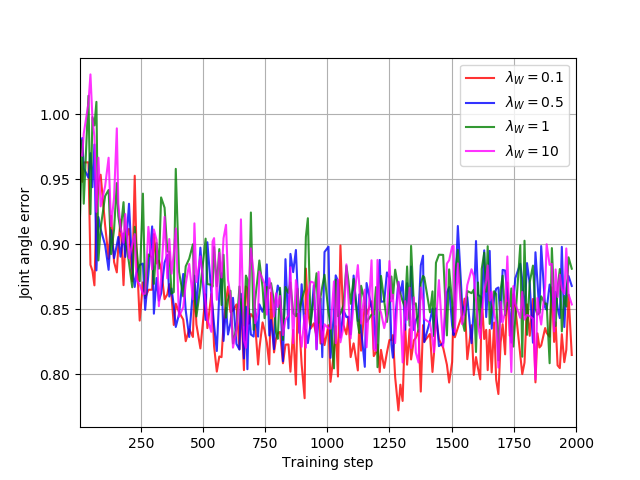}}
  \hfill
  \subfigure[$L_{pos}$ with respect to $\lambda_W$]
  {\includegraphics[width=0.33\linewidth]{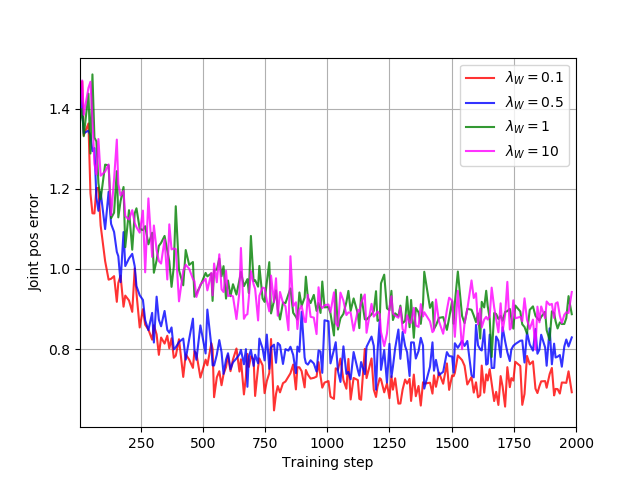}}
  \hfill
  \subfigure[$L_{M}$ with respect to $\lambda_W$]
  {\includegraphics[width=0.33\linewidth]{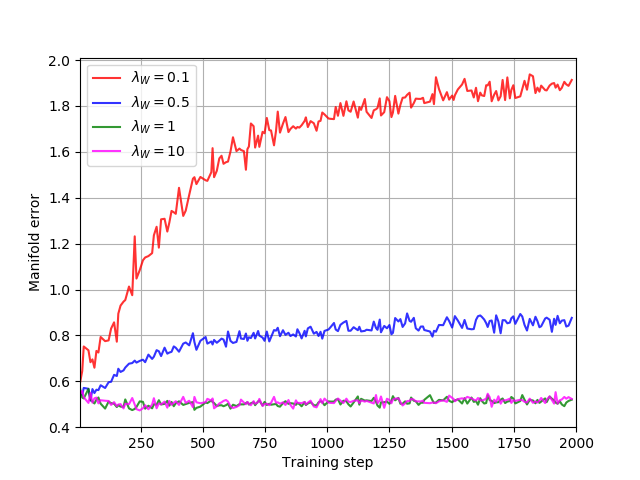}}
  \hfill
\caption{Reconstruction errors of joint angle, joint position and manifold according to training step while adjusting $\lambda_W$ for H3.6M dataset.}
\label{fig:tradeoff}
\end{figure*}

This experiment examines the effect of different settings of the weight $\lambda_W$ for the regularization  on the reconstruction errors ($L_{ang}$ and $L_{pos}$) and on motion manifold ($L_M$) on the test set. We employed {\mD} model for this experiment to exclude the effect of other loss terms. 
Figure~\ref{fig:tradeoff} (a) and (b) show that the joint reconstruction errors decrease as $\lambda_W$ becomes smaller, which makes {\mD} model closer to a pure autoencoder, sacrificing the ability to enforce a prior over the motion manifold space while obtaining better reconstruction loss. For the same reason, Fig.~\ref{fig:tradeoff} (c) shows that the motion manifold reconstruction error $L_M$ decreases as $\lambda_W$ becomes larger. 

As our goal is to obtain an effective motion manifold that is able to generate realistic motions, it is important to find a suitable set of weight parameters that compromise among different qualities.

{\green 
\subsubsection{Adversarial loss and explicit feeding manifold vector}
\label{subsec:ZAd}
\begin{figure}[h]
  \centering
  \includegraphics[width=0.9\columnwidth]{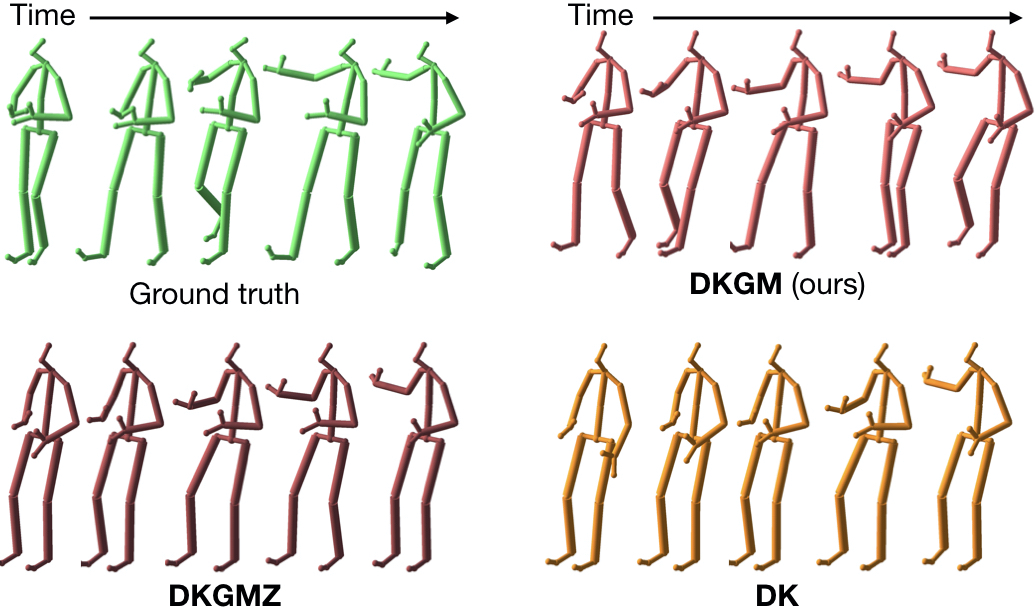}
  \caption{{\green Reconstruction results of different loss combinations for a posing while walking motion. Supplementary video includes full motions.} }
  \label{fig:ablation}
\end{figure}

Here we discuss the effects of adversarial loss (Sec.~\ref{par:ad_loss}) and explicitly feeding motion manifold vector to the decoders on motion quality.
First, Table~\ref{table:H3.6M_general} shows that {\mDKG} decreases $E_r$ from {\mDK} only slightly. However, Fig.~\ref{fig:ablation} shows that {\mDK} cannot properly reconstruct the original motion,
reconstructing only posing motion from the original motion of posing with walking.
In contrast, {\mDKG} improves the overall motion quality by better reconstructing the behaviors in the original motion.
Comparing our method ({\mDKGM}) and {\mDKGMZ}, the latter results in lower $E_r$ and $E_p$ than our method as shown in Table.~\ref{table:H3.6M_general}. 
However, Fig.~\ref{fig:ablation} reveals that {\mDKGMZ} fails to capture walking motion. We conjecture that  directly feeding manifold vector to decoder reduces reconstruction loss by explicitly retaining the motion manifold vector, but tends to converge to mean pose.
In contrast, our method successfully reconstructs the original posing with walking behavior.
This observation suggests that, while the joint reconstruction error is an important indicator of  motion quality, it may not appropriately assess the motion quality in terms of reconstructing the original behaviors.

}

\subsection{Random motion samples}
\label{subsec:sampling}
To verify whether the latent motion manifold can create meaningful motions, we randomly sampled $P_Z$ and decoded to obtain motions. 
We extracted 30 random samples from the motion manifold learned with H3.6M dataset. 
\begin{figure}[h]
  \centering
  \includegraphics[width=0.99\linewidth]{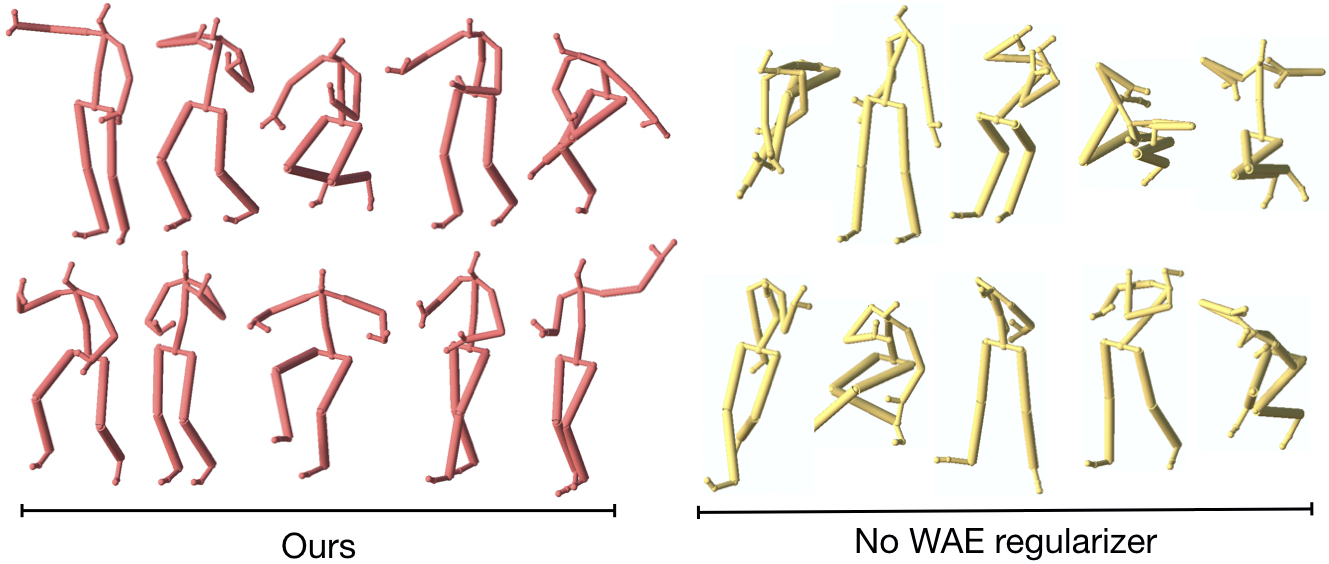}
  \caption{Results of randomly sampling motions from the motion manifold $P_Z$.}
  \label{fig:randomsampling}
\end{figure}
Figure \ref{fig:randomsampling} is the results of random sampling from $P_Z$, and one can see that our method can create various actions including sitting, crossing the legs, and resting on the wall. This result suggests that our motion manifold and decoder can create a wide range of plausible behaviors.

{\green
To examine the importance of WAE, we experimented random sampling by replacing the WAE regularizer with a simple $L_2$-norm ${\lVert z \rVert}^2$ loss. 
Sampled motions from this method, as shown in Fig. \ref{fig:randomsampling} (right), often show unnatural poses and extreme joint rotations. 
This experiment shows that the WAE regularizer not only helps achieve the desired motion manifold distribution but also improves quality of motion sampling.
}

\subsection{Motion interpolation with latent motion manifold}
\label{subsec:interpolation}
We can interpolate two different motions by encoding them into the latent motion manifold and then performing linear interpolation between the encoded motion manifold vectors.
The resulting interpolated motion created by our method is not just frame-by-frame interpolation, but may contain meaningful transition between the input motions.
For example, interpolating sitting down motion and photo taking motion creates hand raising motion to prepare to take a picture from sitting posture. 
When waiting and smoking motions are interpolated, an interesting motion that a character seems tired of waiting and starts to smoke is created. 
The capability of creating such meaningful motions is due to the Wasserstein regularizer that shortens the distance between the encoded vectors by matching the motion manifold to the multivariate normal prior.
Figure \ref{fig:interpolation} and the supplemental video show the interpolated motions.

Figure~\ref{fig:inter_compare} compares our model with \cite{holden2015learning} with respect to interpolation. See supplementary material for the implementation of \cite{holden2015learning}. 
For the interpolation from sitting to walking (top) and from sitting down to taking photo (bottom), our model shows a natural transition between two motions while \cite{holden2015learning} creates somewhat averaged motion between the two motions. 


\begin{figure}[h]
  \centering
  \includegraphics[width=0.99\linewidth]{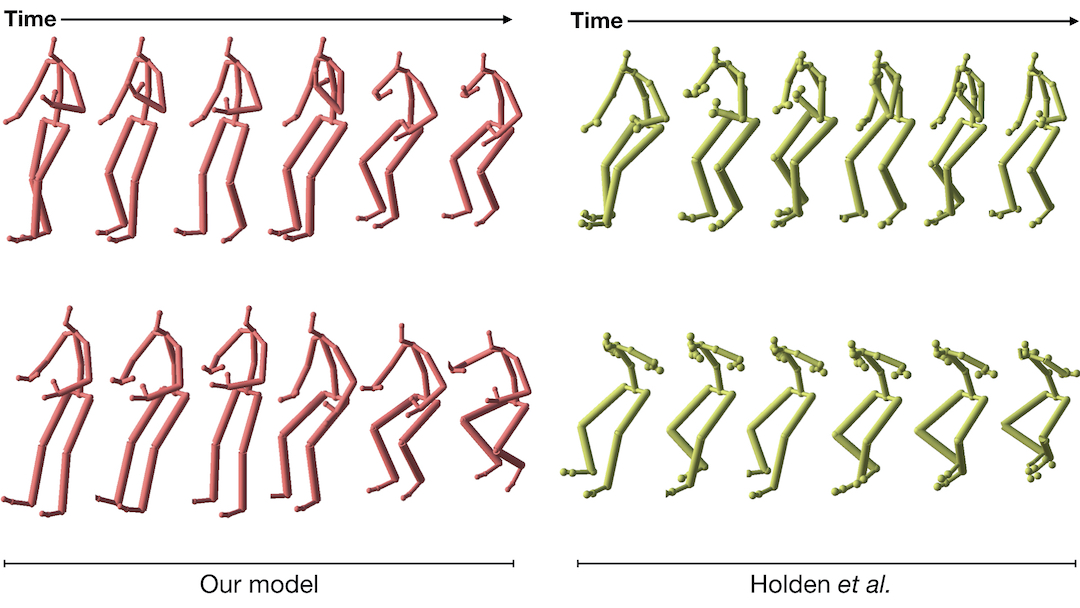}
  \caption{Interpolation from sitting to walking (top) and from sitting down to taking photo (bottom) made by our model (left) and \cite{holden2015learning} (right).}
  \label{fig:inter_compare}
\end{figure}

\subsection{Denoising motion data}
\label{subsec:denoising}
\begin{figure}[h]
  \centering
  \includegraphics[width=0.99\linewidth]{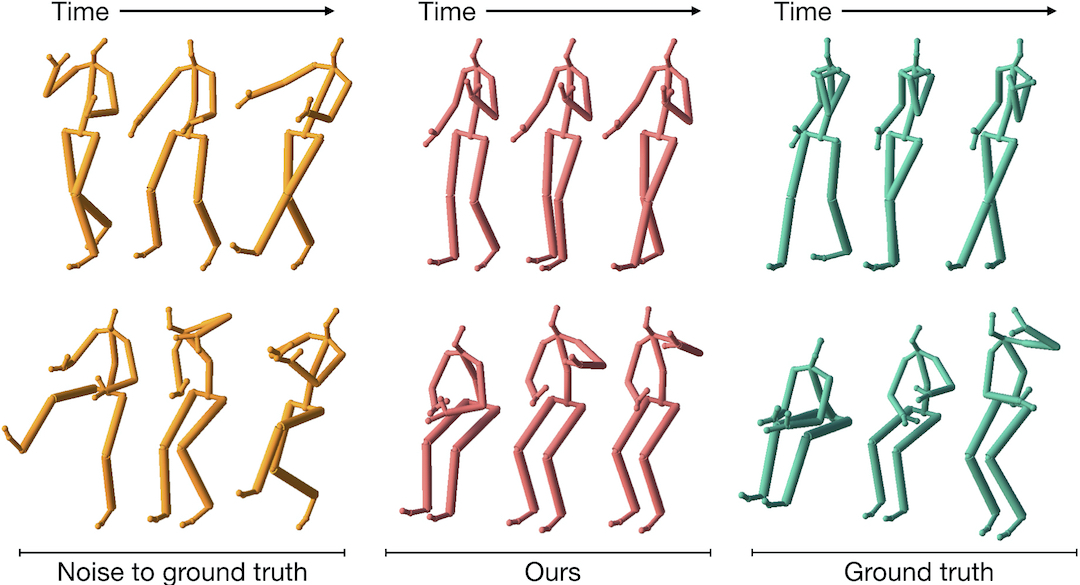}
  \caption{Denoising experiment. Three poses are shown from the noise corrupted motion (orange), denoised motion by our method (coral), and the ground truth motion (green). Two motions (top and bottom) are shown.}
  \label{fig:noise}
\end{figure}

Our motion model can denoise motion data by projecting it to the latent motion manifold and decoding the motion manifold vector to obtain a reconstructed motion.
{\green 
Since the motion manifold is constructed only from human motion capture data, any element in the manifold is likely to be decoded to natural motion. Therefore, denoising effect occurs when noisy motion data is projected to the motion manifold.}
We experiment on the denoising capability of our method in the similar manner as in \cite{holden2015learning}.
We generate noise corrupted motion by randomly setting joint angles to zero with a probability of 0.5, which makes half of the joint angle information meaningless.
Figure \ref{fig:noise} shows the denoised results which are quite similar to the ground truth motions.

\subsection{Motion analogy}
\label{subsec:analogy}
\begin{figure}[h]
\centering
  \subfigure[{Motion analogy among ``Walking with posing'', ``Walking'' and ``Sitting'' actions.}]
  {\includegraphics[width=0.99\linewidth]{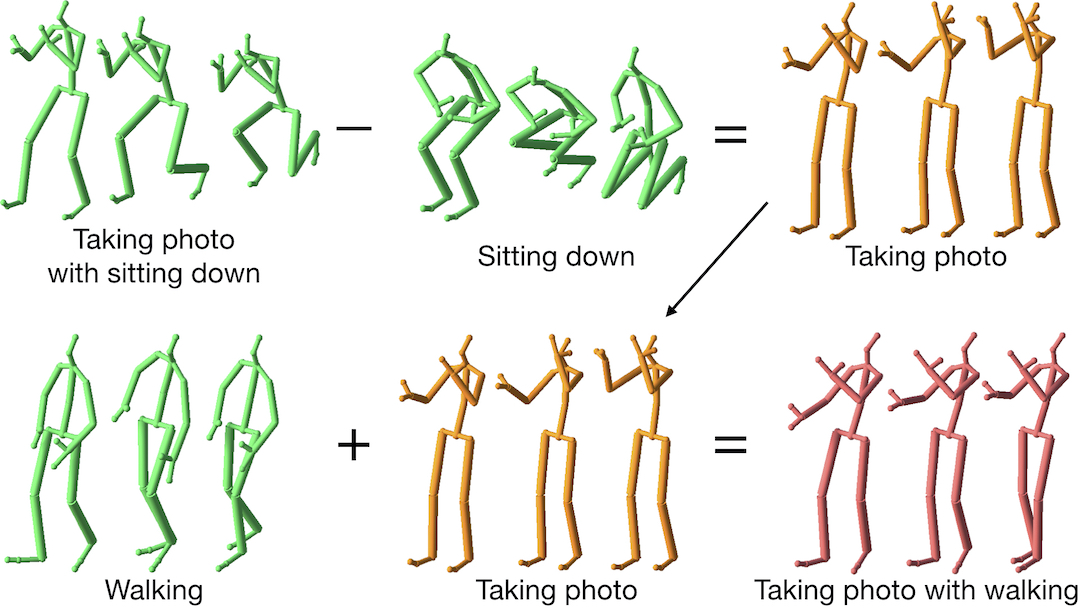}}
  \hfill
  \subfigure[{Motion analogy among  ``Smoking with sitting'', ``Sitting'' and ``Walking'' actions.}]
  {\includegraphics[width=0.99\linewidth]{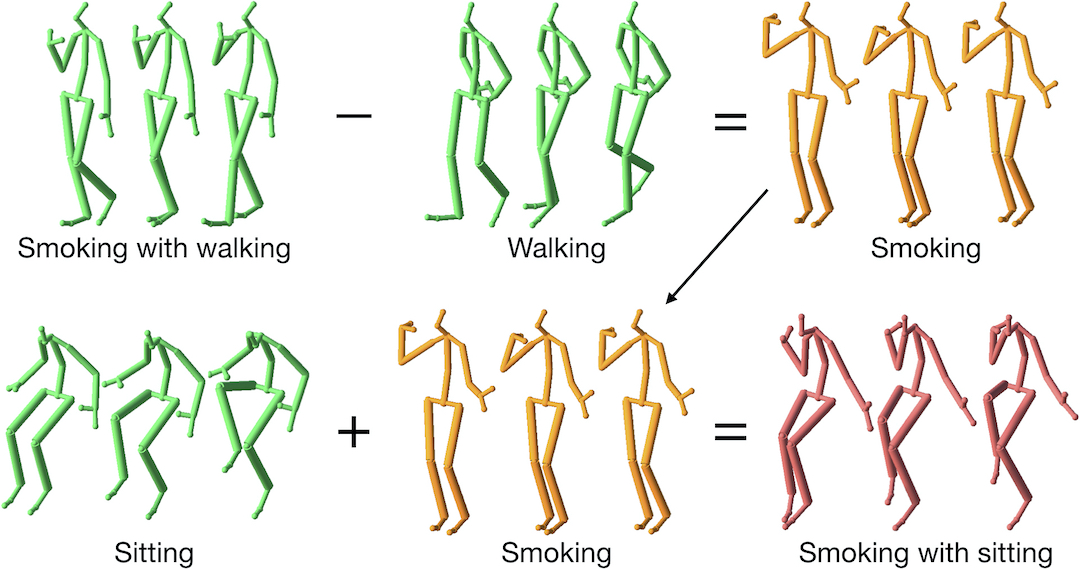}}
  \hfill
\caption{{\green Motion analogy experiments performing arithmetic operations in the motion manifold.}}
\label{fig:analogies}
\end{figure}

Through motion analogy, we can understand how our model organizes motion manifold to represent the feature of actions.
Details about analogy can be found in \cite{white2016sampling}.
We perform vector algebraic operations with the latent vectors encoded from different motions and explore how the model organizes the latent space to represent motions. 
{\green Figure \ref{fig:analogies} (a) shows that subtracting a motion manifold vector for ``sitting down'' motion from ``taking photo with sitting down'' motion creates a vector representing ``taking photo'' motion. The character is standing because a zero vector in our motion manifold corresponds to an idle standing motion. Subsequently, when an encoded ``walking'' motion manifold vector is added, the motion vector becomes a vector for ``taking photo with walking'' motion. 
Figure \ref{fig:analogies} (b) shows a similar analogy among ``walking'', ``smoking with walking'',  and ``sitting'' motions.}

\begin{figure}[h]
  \centering
  \includegraphics[width=0.99\linewidth]{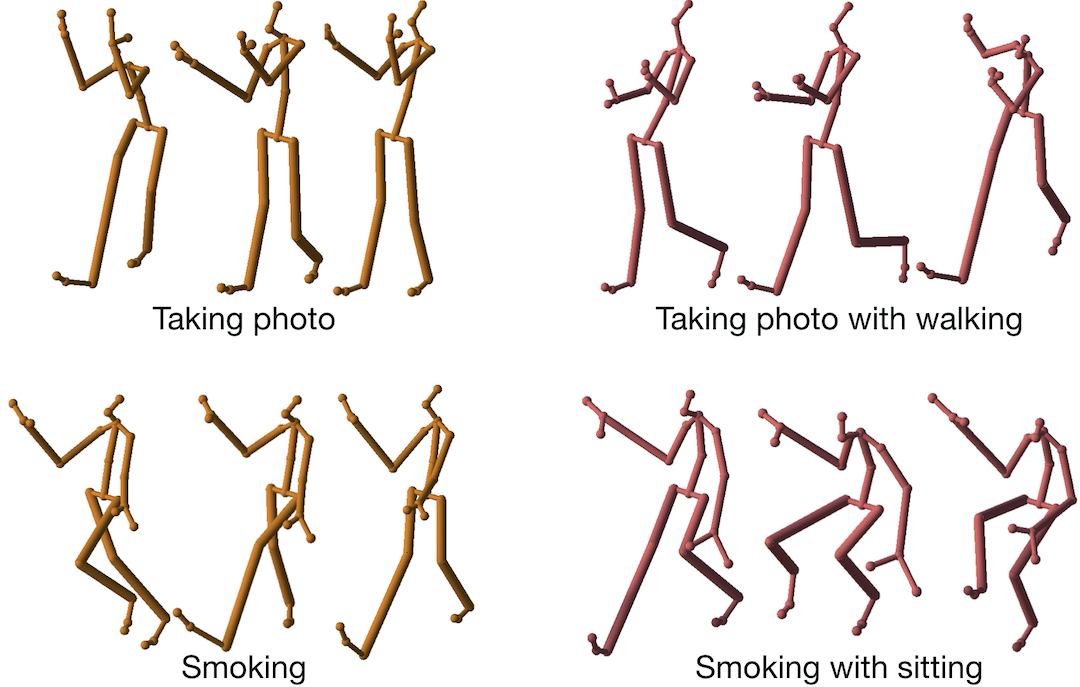}
  \caption{{\green Motion analogy experiment with \cite{holden2015learning}.}}
  \label{fig:anal_Holden}
\end{figure}

Figure \ref{fig:anal_Holden} shows the experiments of performing analogy with \cite{holden2015learning}.
{\green Figure \ref{fig:anal_Holden} (top) is the result of taking photo (left) and taking photo with walking (right) that correspond to Fig. \ref{fig:analogies} (a), and Fig. \ref{fig:anal_Holden} (bottom) shows smoking and smoking with sitting to compare with Fig. \ref{fig:analogies} (b).} One can see that the motion manifold obtained with \cite{holden2015learning} does not support analogy on the motion manifold.

\section{Conclusion and future work}
\label{sec:conclusion}
In this paper, we presented a novel sequential network for constructing a latent motion manifold for modeling human motion.
The main contributions of our method are the combined decoder for the joint rotation and joint velocity, and considering both the joint rotations and positions by adding the FK layer in both decoders, which improve the reconstruction accuracy.
In addition, we composed a set of loss functions, each of which contribute to enhancing the quality of  motions generated from the motion manifold space from different aspects. 
The capabilities of our model have been examined through various experiments such as random sampling, motion interpolation, denoising, and motion analogy.

Our  method  has  several  limitations.
First, as a sequence-to-sequence framework, the performance of our model degrades if trained to produce motions longer than 10 seconds. 
{\green 
The supplementary video shows randomly generated motions with our network being trained to learn 300 frames (approx. 13 seconds). Resulting motions tend to lose details.
}
This limitation may be alleviated by employing an attention mechanism ~\cite{luong2015effective, bahdanau2014neural}.
Second, the encoded motions tend to be smoothed in the process of matching the latent motion manifold to the prior distribution through the regularizer.
For example, motions that contain frequent hand shaking, such as ``walking with dog'' or ``discussion'' motions in H3.6M dataset, lose fine details when reconstructed. Overcoming these limitations will be important future work. 

{\green We only considered joint rotations in the encoder, but incorporating additional information, such as joint positions and velocities, may be beneficial to achieve better motion qualities.
In addition, in the process of learning a motion manifold, loss terms to check validity of motions, such as joint limit, velocity limit and foot sliding, are not needed as all input motion data are considered valid. However, when an actual motion is sampled from the manifold and applied to an environment, such criteria may need to be checked. 
}

Most studies on motion space learning have focused on representing a wide range of motion categories with a compact representation. In fact, the range of motion categories is only one aspect of the variedness of human motions. 
Even a single motion category such as walking exhibits widely different styles depending on gender, body scale, emotion, and personality. Developing a motion manifold that can generate stylistic variations of motion is another important future research direction.

\section*{Acknowledgement}
This work was supported by Giga Korea Project (GK17P0200)
and Basic Science Research Program (NRF-2020R1A2C2011541) funded
by Ministry of Science and ICT, Korea.

\bibliographystyle{eg-alpha-doi} 
\bibliography{references}       



\onecolumn
\renewcommand*\appendixpagename{Supplementary Material}
\renewcommand*\appendixtocname{Supplementary Material}

\begin{appendices}
\section{Network structures and experimental setup}
\label{sec:appendix}
The following will describe details of the network structures of the models used for comparison and the hyperparameters of each model.
For all models, we use batch size of 30 for H3.6M data set. The number of training epochs is 500. Dimension of motion manifold is 64. 


\subsection{Sequence-to-sequence model}
The seq2seq model used in our comparison has a similar structure as that in \cite{martinez2017human}. The only difference is that a fully connected layer for motion manifold generation exists between the motion manifold and the encoder/decoder.
The encoder and decoder are implemented with a 1-layer Gated Recurrent Unit (GRU) with 1024 dimensional hidden state.
The decoder includes a residual network.
\begin{equation}
    \begin{aligned}
    &\, Enc : \,\, \mQ_{t:(t+\Delta t -1)} \in \mathcal{Q} \rightarrow \text{GRU}_{1024} \rightarrow \text{FC}_{64}
    \rightarrow Z \in \mathcal{Z} \\
    &\, Dec: \,\, Z \in \mathcal{Z} \rightarrow \text{FC}_{1024} \rightarrow \text{Res}\left[\text{GRU}_{1024} \rightarrow \text{FC}_{51}\right]
    \xrightarrow{\text{Flip}} \widehat{\mQ}_{t:(t+\Delta t -1)} \in \mathcal{Q},
    \end{aligned}
\end{equation}
where Res and FC denote the residual network and the fully connected layer, respectively.

For training, we use Adam optimizer with a learning rate of 0.001 and decaying rate of 0.999 per training step, and clip the gradients to scale 1.
The loss function is the same as our motion reconstruction loss.

\subsection{Convolutional model}
Convolutional models have the same structure as \cite{holden2015learning, holden2016deep}.
Unlike our model and seq2seq model, the convolutional model used joint position $\mP$ for training.
Both encoder and decoder use one 1D convolutional layer with a temporal filter of width 15 and the number of hidden units being 256.
The encoder passes $\mP_{t:(t+\Delta t -1)}$ through the convolutional layer, max pooling, ReLU, and finally dropout to map to the motion manifold space.
Decoder generates $\widehat{\mP}_{t:(t+\Delta t -1)}$ by passing through the inverse convolutional layer after upsampling the motion manifold vector $Z$.
\begin{equation}
    \begin{aligned}
    &\, Enc : \,\, \mP_{t:(t+\Delta t -1)} \in \mathcal{Q} \rightarrow \text{Conv}_{256} \rightarrow
    \text{MaxPool} \rightarrow \text{ReLU} \rightarrow \text{Dropout}
    \rightarrow Z \in \mathcal{Z} \\
    &\, Dec: \,\, Z \in \mathcal{Z} \rightarrow \text{UpSampling} \rightarrow
    \text{InConv}_{256}
    \rightarrow \widehat{\mP}_{t:(t+\Delta t -1)} \in \mathcal{Q},
    \end{aligned}
\end{equation}
We use Adam optimizer with learning rate of 0.001 and decay learning rate of 0.999.
Loss function is the mean squared error of the reconstructed motion $\widehat{\mP}_{t:(t+\Delta t -1)}$ and the ground truth motion $\mP_{t:(t+\Delta t -1)}$.

\subsection{Our model}
Our encoder and decoder architectures are detailed in Sec.~\ref{sec:method}. Here we describe the network weights and parameters for the training.
Both the encoder and decoder include one layer of GRU with 1024 cell size and dropout of 0.2.
Our model has three fully connected layers: $\text{FC}_{64}$ from the encoder to the motion manifold, $\text{FC}_{1024}$ from the motion manifold to the decoder, and $\text{FC}_{51}$ from the decoder to the output motion.

For the discriminator network used for the adversarial loss, we use a total of four 1D convolutional layers.
First three layers (Layers 1-3) have the kernel size 4, stride 2, 1 reflect padding, and leakyReLU activation with leak of 0.2. Layer 4 has kernel size 1, stride 1, and Relu activation.
The number of units in Layers 1 and 4 are 32 and 1, respectively. 
Layers 2 and 3 have 64 and 128 units respectively with batch normalization.  
For training, we use Adam optimizer with a learning rate of 0.001 for both the motion manifold networks and discriminator, and clip the GRU gradients by the global norm of 1. 



\end{appendices}

\end{document}